%% file: CFS03_18.tex
\documentclass[aps,prd]{revtex4}
\usepackage{epsf}
\usepackage{epsfig}
\usepackage{epstopdf}
\usepackage{graphicx}
\usepackage{subfigure}
\usepackage{amssymb}
\usepackage{color}
\newcommand{\be}{\begin{equation}}
\newcommand{\ee}{\end{equation}}
\newcommand{\ba}{\begin{eqnarray}}
\newcommand{\ea}{\end{eqnarray}}
\newcommand{\no}{\nonumber\\}

\newcommand{\grts}{\raise.3ex\hbox{$>$\kern-.75em\lower1ex\hbox{$\sim$}}}
\newcommand{\lets}{\raise.3ex\hbox{$<$\kern-.75em\lower1ex\hbox{$\sim$}}}

\begin{document}
\title{The Next-to-Minimal Two Higgs Doublet Model}
\author{Chien-Yi Chen}\email{cychen@bnl.gov}
\affiliation {Department of Physics, Brookhaven National Laboratory, Upton, New York, 11973}

\author{Michael Freid}\email{mcfreid@email.wm.edu}
\affiliation {High Energy Theory Group, College of William and Mary, Williamsburg, Virginia 23187, U.S.A.\\  Thomas Jefferson National Accelerator Facility, Newport News, Virginia 23606}

\author{Marc Sher}\email{mtsher@wm.edu}
\affiliation {High Energy Theory Group, College of William and Mary, Williamsburg, Virginia 23187, U.S.A}

\vskip 0.5cm
\date{\today}

\begin{abstract}
The simplest extension of the Two Higgs Doublet Model is the addition of a real scalar singlet, $S$.      The effects of mixing between the singlet and the doublets can be manifested in two ways.   It can modify the couplings of the $126$ GeV Higgs boson, $h$, and it can lead to direct detection of the heavy Higgs at the LHC.   In this paper, we show that in the type-I model, for heavy Higgs masses in the $200-600$ GeV range, the latter effect will be detected earlier than the former for most of parameter space.   Should no such Higgs be discovered in this mass range, then the upper limit on the mixing will be sufficiently strong such that there will be no significant effects on the couplings of the $h$ for most of parameter space.   The reverse is true in the type-II model, the limits from measurements of the couplings of the $h$ will dominate over the limits from non-observation of the heavy Higgs.

\end{abstract}

\maketitle

\section{Introduction}

Following the discovery of the Higgs boson\cite{atlas,cms}, there has been interest in finding constraints on extensions of the Standard Model from measurements of the Higgs gauge and Yukawa couplings as well as from heavy Higgs searches.    One of the most popular extensions of the Standard Model (SM) is the Two Higgs Doublet Model (2HDM)\cite{Branco:2011iw}.    Even before the formal announcement of the discovery, there were some studies\cite{Ferreira:2011aa,Ferreira:2012my} of the bounds on the 2HDM parameter space from the preliminary LHC data, followed by a flurry of papers\cite{
Barger:2013mga,Drozd:2012vf,Chang:2012ve,Alves:2012ez,Craig:2012vn,Craig:2012pu,Bai:2012ex,Azatov:2012qz,Dobrescu:2012td,Ferreira:2012nv,Krawczyk:2013gia,Coleppa:2013dya,Altmannshofer:2012ar,Basso:2012st,Chang:2012zf,Chen:2013kt,Celis:2013rcs,Grinstein:2013npa,Craig:2013hca,Chiang:2013ixa,Barroso:2013zxa,Chen:2013qda,Eberhardt:2013uba,Chen:2013rba,Barger:2013ofa,Belanger:2013xza,Chpoi:2013wga}
 just following the announcement.   Often these papers have examined constraints not only for the current data set, but for the projected bounds once a certain luminosity at 14 TeV has been reached.
 
 In this paper, we will consider the simplest extension of the 2HDM, or the next-to-minimal 2HDM (N2HDM), in which a real singlet is added.   For simplicity, we will assume a $Z_2$ symmetry in which $S\rightarrow -S$, but will discuss relaxing this assumption later.   If there is no mixing between the singlet and the two Higgs doublets, then the singlet would not interact with any SM fields and would thus be irrelevant for current LHC experiments.
 In the presence of such mixing, however, one can expect two effects.   
 
 The first effect is a change in the gauge and Yukawa couplings of the Higgs doublets.   As a simple illustration of this, in all 2HDMs, there is a sum rule in which the sum of the squares of the neutral scalar gauge couplings, ${\cal C}_{hWW}^2 + {\cal C}_{HWW}^2$, where $h$ and $H$ are the mass eigenstates,  is the same as the SM coupling squared.   This sum rule will be violated if there is a singlet which mixes with the doublets.    Thus, one expects projected bounds from 2HDM studies to be altered.   The second effect is that mixing will allow for direct detection of the heavy Higgs (which is mostly singlet), through, for example, $gg \rightarrow S \rightarrow ZZ$.    This should show up in heavy Higgs searches at ATLAS and CMS.       We will ask the following question:  As more data are collected by the LHC, which of these two effects (modifications of the 2HDM couplings or heavy Higgs direct detection) will appear first?
 
 Recently, Cheung, et al.\cite{Cheung:2013bn} asked a similar question.   Although they did consider models with two doublets and a singlet,   their work was much more general, including many singlets, supersymmetric models, etc.   They asked whether precision Higgs measurements would be the first signature of these models.   Our work will be more focused, concentrating only on the 2HDM models with a singlet added. We will also concentrate on direct detection.    Other papers \cite{Dawson:2009yx,Chang:2012ta,Gupta:2013zza} have studied models with a single Higgs doublet and an additional singlet.
 
 \section{The Model}

Models with two Higgs doublets generally have tree level flavor changing neutral currents.   To avoid these, one must couple all quarks of a given charge to a single Higgs doublet.   This is generally accomplished by imposing a $Z_2$ symmetry (which is softly broken).   This can be done in four ways, leading to the familiar four 2HDM's:  type-I, type-II, Lepton Specific and Flipped.   See Ref. \cite{Branco:2011iw} for a review.   The gauge and Yukawa couplings of the light Higgs, $h$, are given in Table I.   Here, $\tan\beta$ is the ratio of vacuum expectation values $v_2/v_1$ and $\alpha$ is the mixing angle which diagonalizes the neutral scalar mass matrix.
\begin{table}[t]
\caption{Light Neutral Higgs ($h$) Couplings in the 2HDMs}
\centering
\begin{tabular}[t]{|c|c|c|c|c|}
\hline\hline
& I& II& Lepton Specific& Flipped\\
\hline
$g_{hVV}$ & $\sin(\beta-\alpha)$ & $\sin (\beta-\alpha)$ &$\sin (\beta-\alpha)$&$\sin (\beta-\alpha)$\\
$g_{ht\overline{t}}$&${\cos\alpha\over\sin\beta}$&${\cos\alpha\over\sin\beta}$&${\cos\alpha\over\sin\beta}$&${\cos\alpha\over\sin\beta}$\\
$g_{hb{\overline b}}$ &${\cos\alpha\over\sin\beta}$&$-{\sin\alpha\over\cos\beta}$&${\cos\alpha\over\sin\beta}$&$-{\sin\alpha\over \cos\beta}$\\
$g_{h\tau^+\tau^-}$&${\cos\alpha\over \sin\beta}$&$-{\sin\alpha\over \cos\beta}$&$-{\sin\alpha\over \cos\beta}$&${\cos\alpha\over\sin\beta}$\\
\hline
\end{tabular}
\label{table:coups}
\end{table}
In this work, the Lepton-Specific (Flipped) will not give appreciably different results than the type-I (type-II) models, and thus we will not discuss them further.

We now add a real singlet, $S$, with a discrete $S \rightarrow -S$ symmetry and will arrange parameters so that $S$ acquires a vacuum expectation value (VEV).     The implications of relaxing this discrete symmetry will be discussed in the section with our  conclusions.    We are not including a complex singlet.  If we did so, then the discrete $Z_2$ symmetry would be promoted to a global $U(1)$, and the spontaneous breaking would lead to a massless pseudoscalar (although this might be phenomenologically acceptable if it does not couple to SM particles).  The Lagrangian is given by

\begin{eqnarray}
V &=&
m^2_{11}\, \Phi_1^\dagger \Phi_1
+ m^2_{22}\, \Phi_2^\dagger \Phi_2 -
 \mu^2\, \left(\Phi_1^\dagger \Phi_2 + \Phi_2^\dagger \Phi_1\right) + \frac{1}{2}m^2_S S^2
+ \frac{\lambda_1}{2} \left( \Phi_1^\dagger \Phi_1 \right)^2
+ \frac{\lambda_2}{2} \left( \Phi_2^\dagger \Phi_2 \right)^2
\no & &
+ \lambda_3\, \Phi_1^\dagger \Phi_1\, \Phi_2^\dagger \Phi_2
+ \lambda_4\, \Phi_1^\dagger \Phi_2\, \Phi_2^\dagger \Phi_1
+ \frac{\lambda_5}{2} \left[
\left( \Phi_1^\dagger\Phi_2 \right)^2
+ \left( \Phi_2^\dagger\Phi_1 \right)^2 \right]
\no & &
+\frac{1}{8} \lambda_6 S^4 +\frac{1}{2} \lambda_7 \left( \Phi_1^\dagger \Phi_1 \right) S^2
+\frac{1}{2}\lambda_8 \left( \Phi_2^\dagger \Phi_2 \right) S^2,
\end{eqnarray}
where $\Phi_1$ and  $\Phi_2$ are the SU(2) doublet Higgs fields and their VEVs are denoted as $v_1$ and $v_2$, respectively.
The neutral Higgs mass matrix is now a $3\times 3$ matrix.  We assume that $\lambda_7$ and $\lambda_8$ are small and expand the 
mass matrix to leading order in these couplings.    If $\lambda_7$ and/or $\lambda_8$ are not small, then mixing would be huge and 
one would expect substantial effects on the couplings and production, which we will later see are not allowed.   
With this expansion, we find the gauge and Yukawa couplings of the scalars.  
Note that to zeroth order in $\lambda_7$ and $\lambda_8$, the mass matrix divides into the conventional $2\times 2$ matrix and a $1\times 1$ matrix, 
and thus to this order, $\alpha$ carries the same definition as in the usual 2HDM.   
Since we are only working to leading order in perturbation theory, the expression for the corrections can use this definition as well.

The additional $\lambda_7$ and $\lambda_8$ terms will affect the usual theoretical constraints on the parameters of the Higgs potential 
from vacuum stability and perturbative unitarity.   Looking at the potential at large scales, we can show that the stability bound, 
in addition to $\lambda_1 > 0$, $\lambda_2 > 0$ and $\lambda_6 >0$, is
\begin{equation}
(\lambda_3 + \lambda_4 +\lambda_5)^2 <  \lambda_1 (\lambda_2 - \lambda_8^2/\lambda_6)
\end{equation}
for $\lambda_7 = 0$ and the same with $\lambda_8 \rightarrow \lambda_7, \lambda_1\leftrightarrow\lambda_2$ for $\lambda_8 = 0$.  
If both are nonzero, the expression is much more complicated.   We see that the parameter space is somewhat more constrained 
(requiring $\lambda_8^2 < \lambda_2\lambda_6$, for example if $\lambda_7=0$), but for relatively small $\lambda_7$ and $\lambda_8$, 
these conditions can be satisfied.    Perturbative unitarity bounds are relevant when the couplings become large, generally bounding masses by, 
typically, 600 $-$ 800 GeV.   Since we are dealing with smaller couplings and lower masses, these should not be problematic. 

The results for the gauge couplings are:

\begin{eqnarray}
{\cal C}_{H_1 ZZ} &=&  \frac{v}{4}g^2_Z \cos(\beta-\alpha)\left[ 1 - \frac{1}{2} \Delta^{\prime 2}\cos^2\beta\right]\cr
{\cal C}_{H_2 ZZ} &=&  \frac{v}{4}g^2_Z \sin(\beta-\alpha)\left[ 1 - \frac{1}{2} \Delta^{2}\sin^2\beta\right]\cr
{\cal C}_{H_3 ZZ} &=& - \frac{v}{4}g^2_Z \left[\Delta^\prime\cos\beta\cos(\beta-\alpha) + \Delta\sin\beta\sin(\beta-\alpha)\right]
\label{cv}
\end{eqnarray}

Here, $v^2 = v^2_1 + v^2_2$, $g^2_Z = g^2/\cos^2\theta_W$ and
\begin{eqnarray}
\Delta &\equiv& \frac{\lambda_8 v v_s}{m^2_{H_2} - m^2_{H_3}}\cr
\Delta^\prime &\equiv& \frac{\lambda_7 v v_s}{m^2_{H_1} - m^2_{H_3}}
\label{Del}
\end{eqnarray}
where $v_s$ is the VEV of the singlet.    Note that the fact that $\lambda_7$ and $\lambda_8$ are small does not necessarily imply that
$\Delta$ and $\Delta^\prime$ are small. However, in practice the relevant expansion parameters are $\Delta$ and $\Delta^\prime$, 
and they will typically never exceed 0.35 in this analysis, as we will see below (Table \ref{table:del}, for example.)

For the Yukawa couplings:

\begin{eqnarray}
{\cal C}_{H_1\bar{t}t} &=& \frac{M_t}{v}\frac{\sin\alpha}{\sin\beta}\left[ 1 - \frac{1}{2} \Delta^{\prime 2}\cos^2\beta\right]\cr
{\cal C}_{H_2\bar{t}t} &=& \frac{M_t}{v}\frac{\cos\alpha}{\sin\beta}\left[ 1 - \frac{1}{2} \Delta^{2}\sin^2\beta\right]\cr
{\cal C}_{H_3\bar{t}t} &=& -\frac{M_t}{v}\left[\Delta^\prime\sin\alpha\cot\beta+\Delta\cos\alpha\right]
\label{ct}
\end{eqnarray}
for the top quark.   For the bottom quark in the type-I model, one just replaces $M_t$ with $M_b$.   In the type-II model,
\begin{eqnarray}
{\cal C}_{H_1\bar{b}b} &=& \frac{M_b}{v}\frac{\cos\alpha}{\cos\beta}\left[ 1 - \frac{1}{2} \Delta^{\prime 2}\cos^2\beta\right]\cr
{\cal C}_{H_2\bar{b}b} &=&- \frac{M_b}{v}\frac{\sin\alpha}{\cos\beta}\left[ 1 - \frac{1}{2} \Delta^{2}\sin^2\beta\right]\cr
{\cal C}_{H_3\bar{b}b} &=& -\frac{M_b}{v}\left[\Delta^\prime\cos\alpha-\Delta\sin\alpha\tan\beta\right]
\label{cb}
\end{eqnarray}

In our analysis, we will  ignore $\Delta^\prime$ terms.     They are irrelevant for the gauge and top Yukawa couplings of the light Higgs, $H_2$, and for the singlet couplings they are multiplied by either $\cos(\beta-\alpha)$ or by $\sin\alpha\cot\beta$, both of which are small.   We have checked that inclusion of such terms does not affect our results.

\section{Analysis}

We assume that $H_1$ is the heavier CP even Higgs and $H_2$ the observed Higgs boson, which has a mass at 126 GeV. 
$H_3$ is mostly composed of the singlet state. This can be understood by looking at the couplings in Eqs. 2, 4  and 5
in the limit where $\Delta$ and $\Delta^\prime$ go to zero. It is clear that the couplings of $H_2$ to the SM particles match 
the light neutral Higgs couplings in the 2HDMs, as shown in Table I.  

Using the recent measurements of the signal strengths of the light Higgs 
we perform a $\chi^2$ fit by following the same procedure as described in Refs.\cite{Chen:2013rba,Chen:2013qda}.  
The signal strength is defined as
\begin{equation}
R(i,j) = \frac{\sigma_{\rm prod}(i) \ \text{Br}(j)}{[\sigma_{\rm prod}(i) \ \text{Br}(j)]_{SM}},
\end{equation}
where $\sigma_{\rm prod}(i)$ represents the production cross sections of the Higgs due to the production mechanism $i$, 
such as gluon fusion (ggF), vector boson fusion (VBF), associated production (VH), 
and Higgs production associated with a pair of top quarks (t$\bar{\rm t}$H). Br$(j)$ stands for the branching ratios of the Higgs
decay channels $j$.

$\chi^2$ is defined as follows,
\begin{equation}
\chi^2 = \sum_i {(R_i^{\rm N2HDM}-R_i^{\rm meas})^2\over (\sigma^{\rm meas}_i)^2},
\end{equation}
where $R^{\rm meas}$($\sigma^{\rm meas}$) stands for the central value (uncertainty) of the measured 
signal strength shown in Tables \ref{tab:models1} and \ref{tab:models2} for $H_2$ decaying into
the SM bosons and fermions, respectively. 
$R^{\rm N2HDM}$ denotes the signal strength predicted in the N2HDMs.
When the errors are asymmetric, we have averaged them in quadrature,
$\sigma=\sqrt{{(\sigma_+)^2 +(\sigma_-)^2\over 2}}$.

At 14 TeV, the production rate of Higgs bosons is roughly a factor of three greater than at 8 TeV, and thus after a total luminosity of 300 fb$^{-1}$, one expects the total number of Higgs produced to be increased by a factor of 90 over the number produced in 10 fb$^{-1}$ at 8 TeV.  We will assume that their bound scales as the inverse square root of the number of Higgs produced.  This corresponds to scheme 2 of the CMS high luminosity projections\cite{cmsproj}.   Of course, projecting systematic errors is not easy.    However, since we are using this projection for both the limits on the production as well as errors on the precision measurements, and comparing the two, the precise projections should not significantly affect our conclusions.
We first consider the constraint on $\Delta$ that could be obtained by non-observation of a heavy Higgs in the $H_3 \rightarrow WW,ZZ$ decay mode.    CMS has published bounds based on 10 fb$^{-1}$ of data at 7-8 TeV.   They plot the 95\% confidence limit on the cross section for $pp\rightarrow H_3 \rightarrow WW,ZZ$ as a function of the Higgs mass.   The bound is given in Figure 11 of Ref. \cite{Chatrchyan:2013yoa}.   
From Figure 11 of Ref. \cite{Chatrchyan:2013yoa}, one can see that the upper bound on the cross section, relative to that of the SM, is $0.2$ for masses between $200-450$ GeV, rising to $0.4$ for 600 GeV and to $1.0$ at $700$ GeV.    Thus, we project that at 300 fb$^{-1}$, the upper bound from non-observation will be $0.022 (0.044,0.11)$ for masses between $200-400$ (600, 700) GeV.

How does this translate into a bound on $\Delta$?   Let us give a rough argument, which is later followed by a more detailed analysis.    For all masses between $200$ and $800$ GeV, the dominant decay of a SM Higgs is into $WW$ or $ZZ$.   Even above the top pair threshold, the branching ratio into top quarks never reaches more than $20\%$.   Thus, essentially all Higgs bosons produced, even if the mixing is small, will decay into gauge boson pairs, just as a SM Higgs.   The ratio of $pp\rightarrow H_3 \rightarrow WW,ZZ$ therefore only depends on the production cross section, which primarily (through gluon fusion) depends on the square of the Yukawa coupling to the top quark (with a  correction for large $\tan\beta$ in the type-II model).   This is just $\Delta^2\cos^2\alpha$.   We thus conclude that the bound on $\Delta$ for the heavy Higgs masses between 200 and 450 GeV is
\begin{equation}
\Delta^2\cos^2\alpha < 0.022
\end{equation}
We do not know the value of $\alpha$, of course, but it cannot be too far from zero since we know that both $\sin(\beta-\alpha)$ and $\sin\beta$ are near unity.   Looking at the full allowed parameter space at 300 fb$^{-1}$, one can see that $\cos\alpha > 0.71$.   Combining these gives our result of $\Delta < 0.21$.    This is the upper bound that would be obtained if there is no observation of a heavy Higgs at 300 fb$^{-1}$.    For masses of $600$ GeV, this upper bound increases to $0.29$.    Again, this is very rough, and we will give a more detailed analysis shortly.

We first turn to the precision Higgs coupling measurements to see if there could be any measurable effects of the heavy Higgs.
During the next LHC run, there will be great interest in increasing the precision of Higgs coupling measurements.   Many of the analyses referred to above look at constraints on 2HDM parameters.    Typically, one plots the expected reach for various integrated luminosities as a function of $\cos(\beta-\alpha)$ on one axis and $\tan\beta$ on the other.   Given a point in this plane, the gauge and fermion Yukawa couplings of the 126 GeV Higgs can be determined.   The SM corresponds to $\cos(\beta-\alpha)=0$ and $\tan\beta = \infty$.    Thus, for example, the expected reach of the LHC at 300 fb$^{-1}$ would be represented by a curve in this plane (which includes the SM point).   We now will simply change the gauge and Yukawa couplings to include the effects of $\Delta$, and see how that affects these curves.

\begin{figure}[tb]
\subfigure[]{
      \includegraphics[width=0.36\textwidth,angle=0,clip]{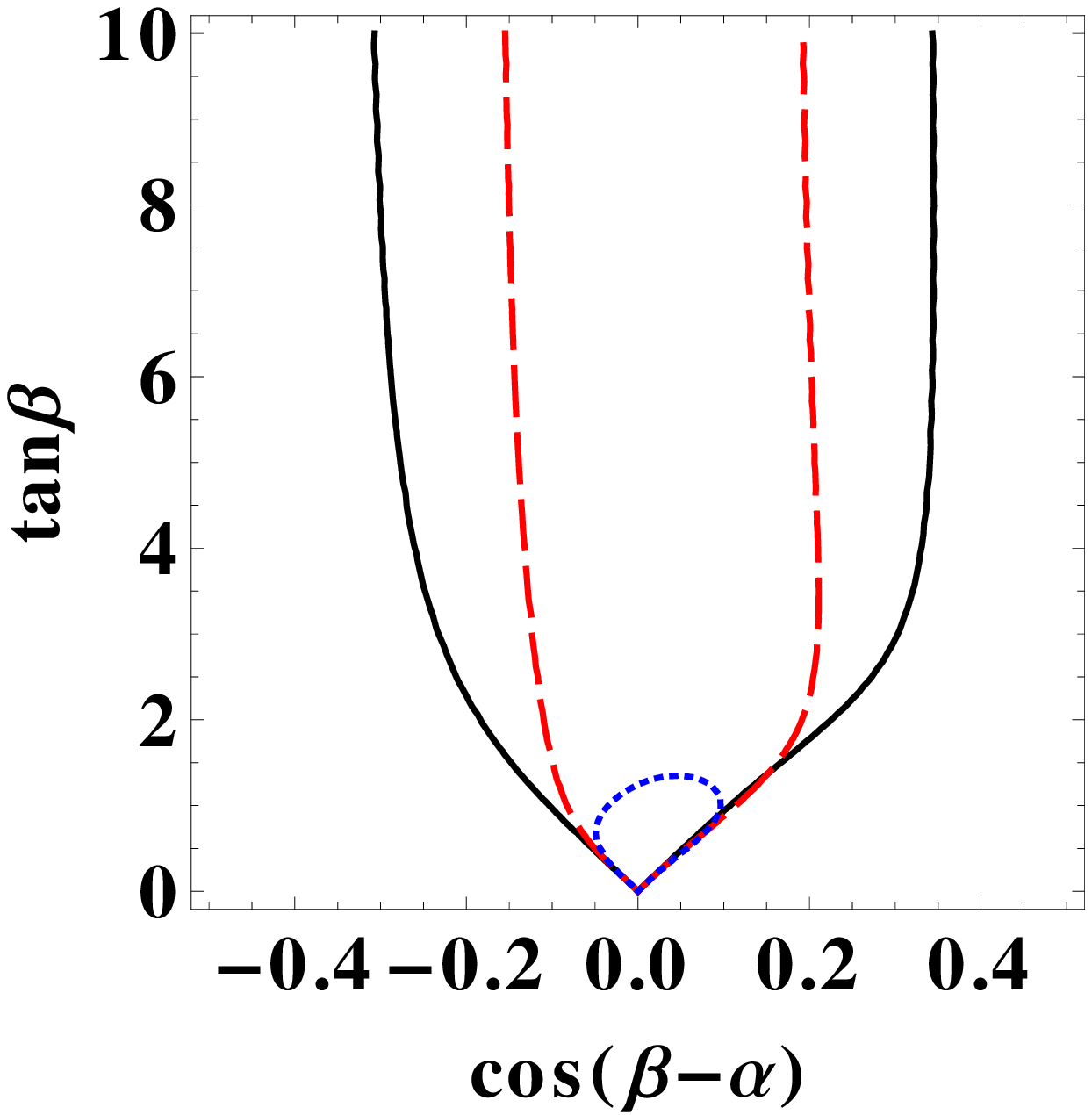}
}
\subfigure[]{
      \includegraphics[width=0.38\textwidth,angle=0,clip]{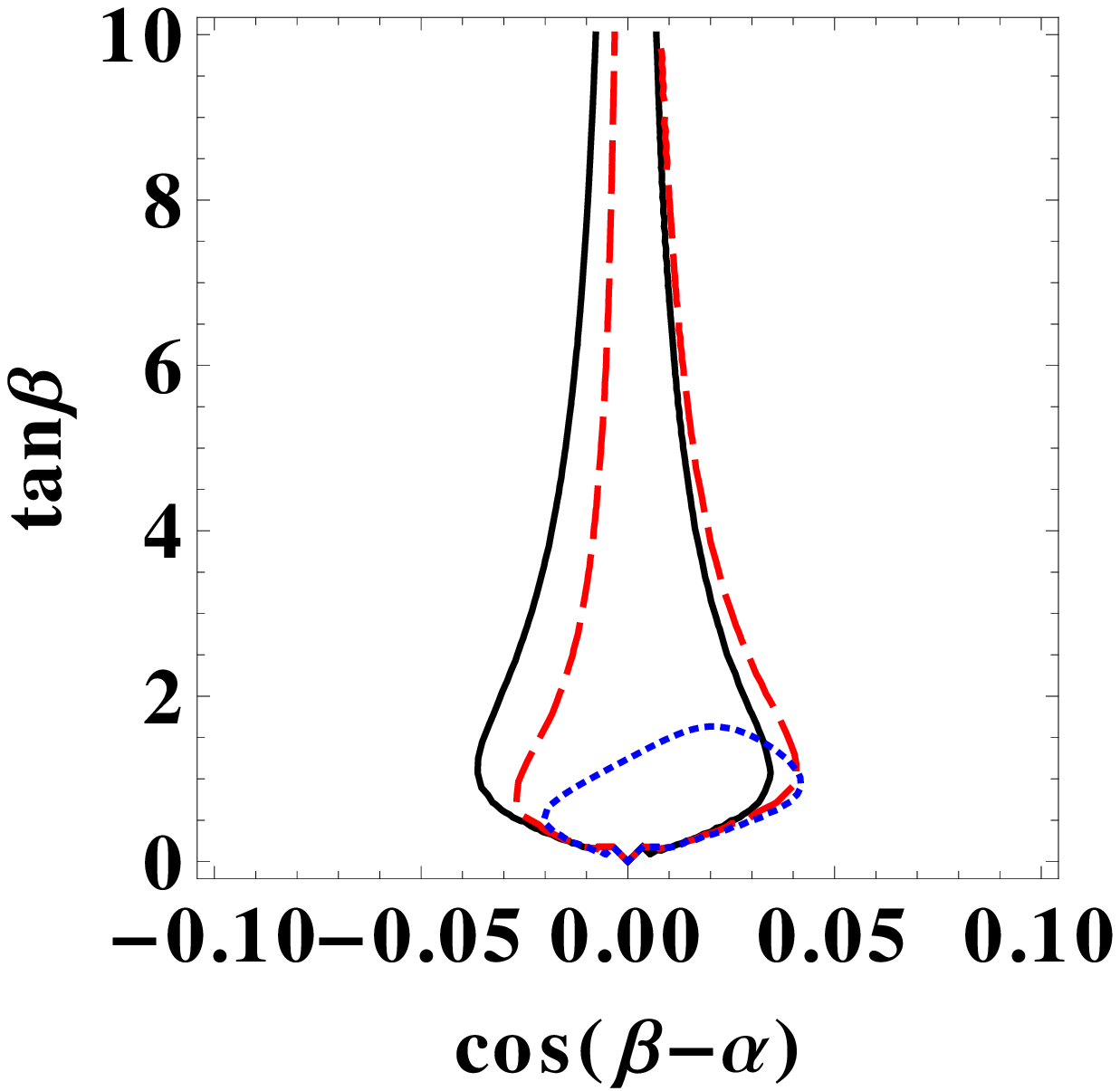}
}
\caption{Allowed regions in the $(\cos(\beta-\alpha),\tan\beta)$ plane  in the type-I (a) and type-II (b)  N2HDMs based on bounds on
$H_2$ obtained by performing a $\chi^2$ analysis.  
The region between the black (solid), red (dashed), and blue (dotted)
 lines is allowed at $95\%$ confidence level corresponding to $\Delta$ = 0, 0.3, and 0.45,  respectively, for an integrated luminosity of 
 300 fb$^{-1}$.
}
\label{vary}
\end{figure}

\begin{figure}[tb]
\subfigure[]{
      \includegraphics[width=0.36\textwidth,angle=0,clip]{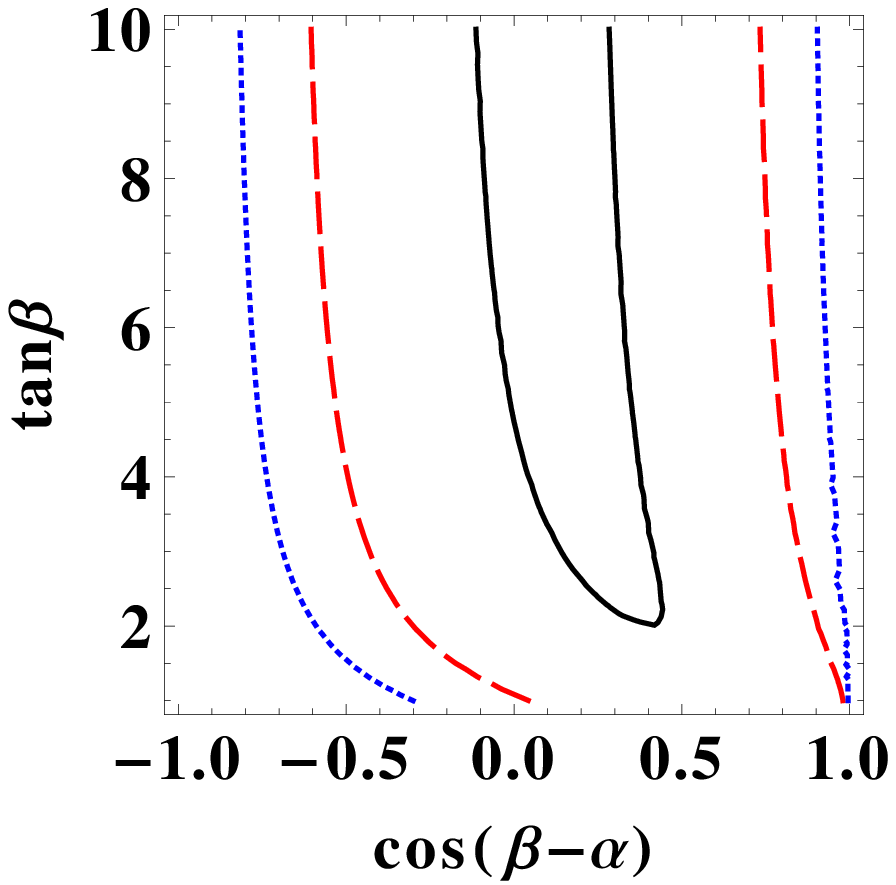}
}
\subfigure[]{
      \includegraphics[width=0.36\textwidth,angle=0,clip]{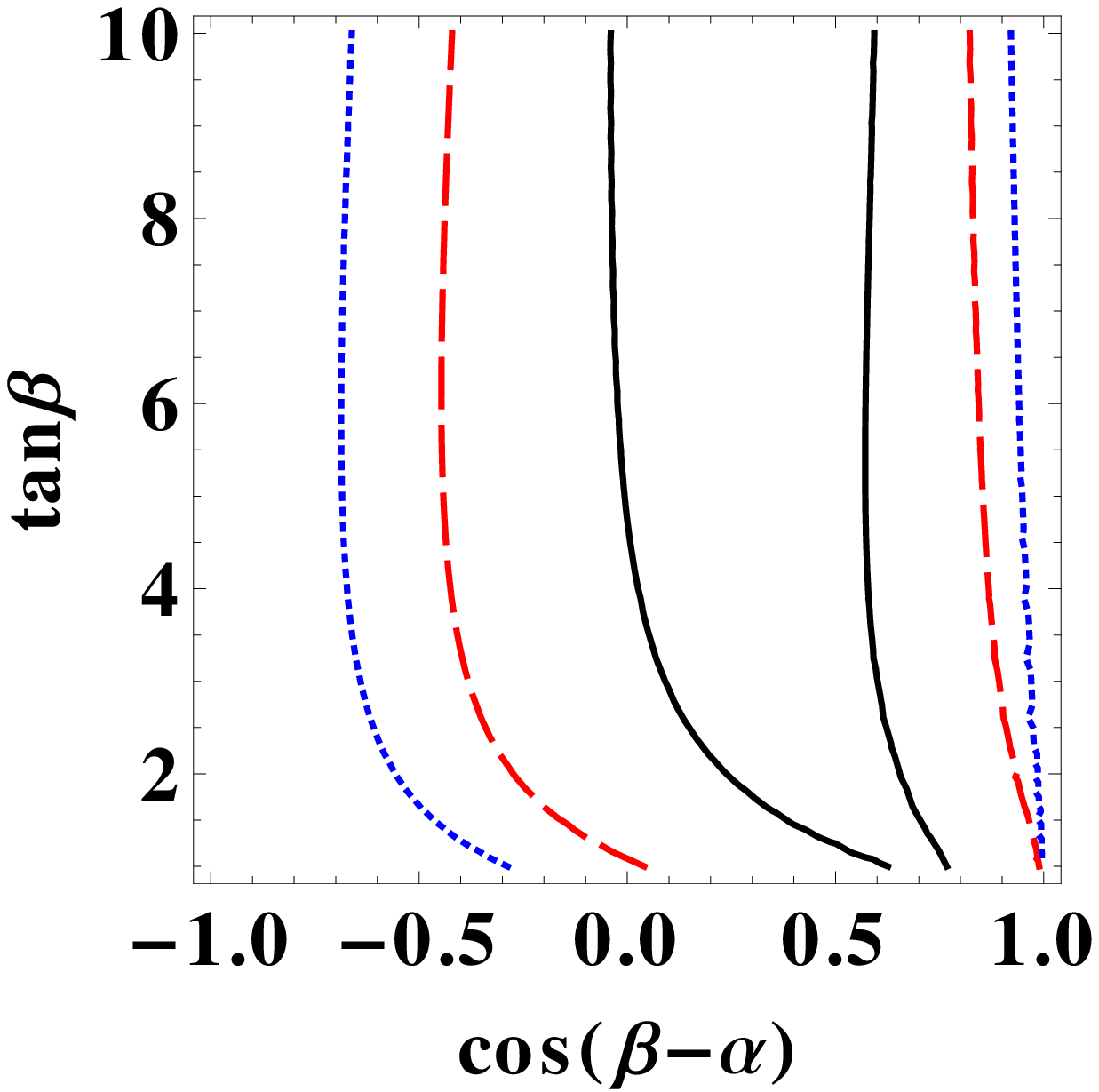}
}
\caption{Allowed regions in the $(\cos(\beta-\alpha),\tan\beta)$ plane in the type-I (a) and type-II (b) N2HDMs based on bounds on
$H_3$ with M$_{H_3}$ = 200 GeV obtained by performing a $\chi^2$ analysis.  
The region outside of the black (solid), red (dashed), and blue (dotted)
 lines is allowed at $95\%$ confidence level corresponding to $\Delta$ = 0.15, 0.2, and 0.3, respectively, for an integrated luminosity of 
 300 fb$^{-1}$.
}
\label{vary-h3}
\end{figure}

As one can see in Eqs. \ref{cv}, \ref{ct}, and \ref{cb}, when $\Delta$ = 0, there is no constraint on $H_3$ since the couplings of $H_3$ to 
the fermions and vector bosons are almost zero (assuming $\Delta^\prime$ is negligible). 
Moreover, since the couplings of $H_1$ to the vector bosons and fermions are not proportional to $\Delta$, 
the constraints on $H_1$ in the N2HDMs are the same as those
in the 2HDMs \cite{Chen:2013rba,Chen:2013qda,Brownson:2013lka,CMS:2013dga}.
The allowed regions in the 
$(\cos(\beta-\alpha),\tan\beta)$ plane in the type-I and type-II N2HDMs are shown in Fig. \ref{vary} based on the projected limits on $H_2$
for an integrated luminosity of 300 fb$^{-1}$.
The region between lines is allowed at 95$\%$ CL.
One can also see that the bound becomes more restrictive with an increase of $\Delta$. For $\Delta$ = 0.45
a large part of the parameter space is excluded for tan$\beta >$  2, in the type-I model. 
We also find that for $\Delta \sim 1$ almost all the parameter space is ruled out at 95$\%$ CL in both N2HDMs.

Should a given point in the $(\cos(\beta-\alpha),\tan\beta)$ plane be realized in Nature, one can then determine the maximum allowed $\Delta$.   For example, in the type-I model (Fig. \ref{vary}a), with $\tan\beta=5$ and $\cos(\beta-\alpha) = 0.1$, one would conclude that there is an upper bound on $\Delta$ which is larger than $0.3$, since it is inside the red line, but smaller than $0.45$, since it is outside the blue line.    A precise analysis for this point gives $\Delta < 0.353$ .

What about the bound from non-observation of the $H_3$?
Fig. \ref{vary-h3} shows the allowed region in the $(\cos(\beta-\alpha),\tan\beta)$ plane in the type-I (a) and type-II (b) N2HDMs 
from the bounds due to non-observation $H_3$ with M$_{H_3}$ = 200 GeV. Note that the region {\it outside} of the black, red, and blue
 lines is allowed at $95\%$ confidence level corresponding to $\Delta$ = 0.15, 0.2, and 0.3, respectively. 
At $\Delta = 0$, there is almost no constraint, but the limits become stronger with the increase of $\Delta$.     Now consider the point in the last paragraph,  a type-I model with $\tan\beta=5$ and $\cos(\beta-\alpha) = 0.1$.   Clearly this will not be acceptable unless $\Delta < 0.15$ (since it is on the black line) -- a precise analysis gives $\Delta <  0.148$  As a result, non-observation (for this particular point) gives a stronger constraint.

\begin{table}[t]
\caption{Benchmark points showing the bounds from the precision measurements of the light Higgs and the non-observation of the 
heavy Higgs with $M_{H_3}=200$ GeV in the type-I model.}
\centering
\begin{tabular}[t]{|c|c|c|c|}
\hline\hline
$\cos(\beta-\alpha)$ & $\tan\beta$ & precision measurements & non-observation\\
\hline
 0.1 & 5  & $\Delta< 0.353$ & $\Delta< 0.148$ \\
-0.1 & 9  & $\Delta< 0.33$ & $\Delta< 0.15$  \\
 0.15 & 5  & $\Delta< 0.33$ & $\Delta< 0.147$ \\
-0.15 & 9  & $\Delta< 0.3$ & $\Delta< 0.152$  \\
 0.2 & 4  & $\Delta< 0.31$ & $\Delta< 0.148$  \\
-0.2 & 4.5  &  $\Delta< 0.23$ & $\Delta< 0.16$  \\
 0.25 & 3  &  $\Delta< 0.25$ & $\Delta< 0.149$  \\
-0.25 & 4  &  $\Delta< 0.1$ & $\Delta< 0.165$  \\
 0.3 &  4 &  $\Delta< 0.16$ & $\Delta< 0.148$  \\
-0.3 &  9 &  $\Delta< 0.05$ & $\Delta< 0.16$  \\
\hline
\end{tabular}
\label{table:del}
\end{table}

Table \ref{table:del} lists a few benchmark points showing the bounds from the precision measurements of the light Higgs and the non-observation of a 
heavy Higgs in the type-I model. The sign of  $\Delta$ can be negative. However, from our analysis we find that flipping the sign of $\Delta$ 
does not change our results. Hence we only show the results for positive $\Delta$.
One can see that non-observation of the heavy Higgs can place a stronger bound compared with the measurement from the
current data. This is true except for the points near $\cos(\beta-\alpha) \simeq |0.3|$, where $\Delta \simeq 0$ from the current measurements.    Thus, except near the edge of the allowed parameter-space, non-observation of the heavy Higgs will give stronger bounds.

We can do this exercise for the entire $\tan\beta,\cos(\beta-\alpha)$ plane.    
In Fig. 3, we have plotted regions of parameter space, for $H_3$ masses of 200 GeV (a) and 600 GeV (b) for an integrated luminosity of 300 fb$^{-1}$.  In the green region, non-observation of the heavy Higgs provides the strongest constraints, while in the red region, precision Higgs measurements are stronger.   The differences are small, but as the heavy Higgs mass goes above 600 GeV, the green region shrinks rapidly.   We see that in the type-I model, non-observation of the heavy Higgs generally provides the strongest constraints on the model. 

For the type-II model, the light Higgs measurements already give very restrictive constraints
compared with those from the non-observation of the heavy Higgs.   Only in the region close to $\cos(\beta-\alpha) \simeq 0$ (corresponding to the SM limit), the non-observation of the heavy Higgs can give a stronger 
limit on $\Delta$.



\begin{figure}[tb]
\subfigure[]{
      \includegraphics[width=0.36\textwidth,angle=0,clip]{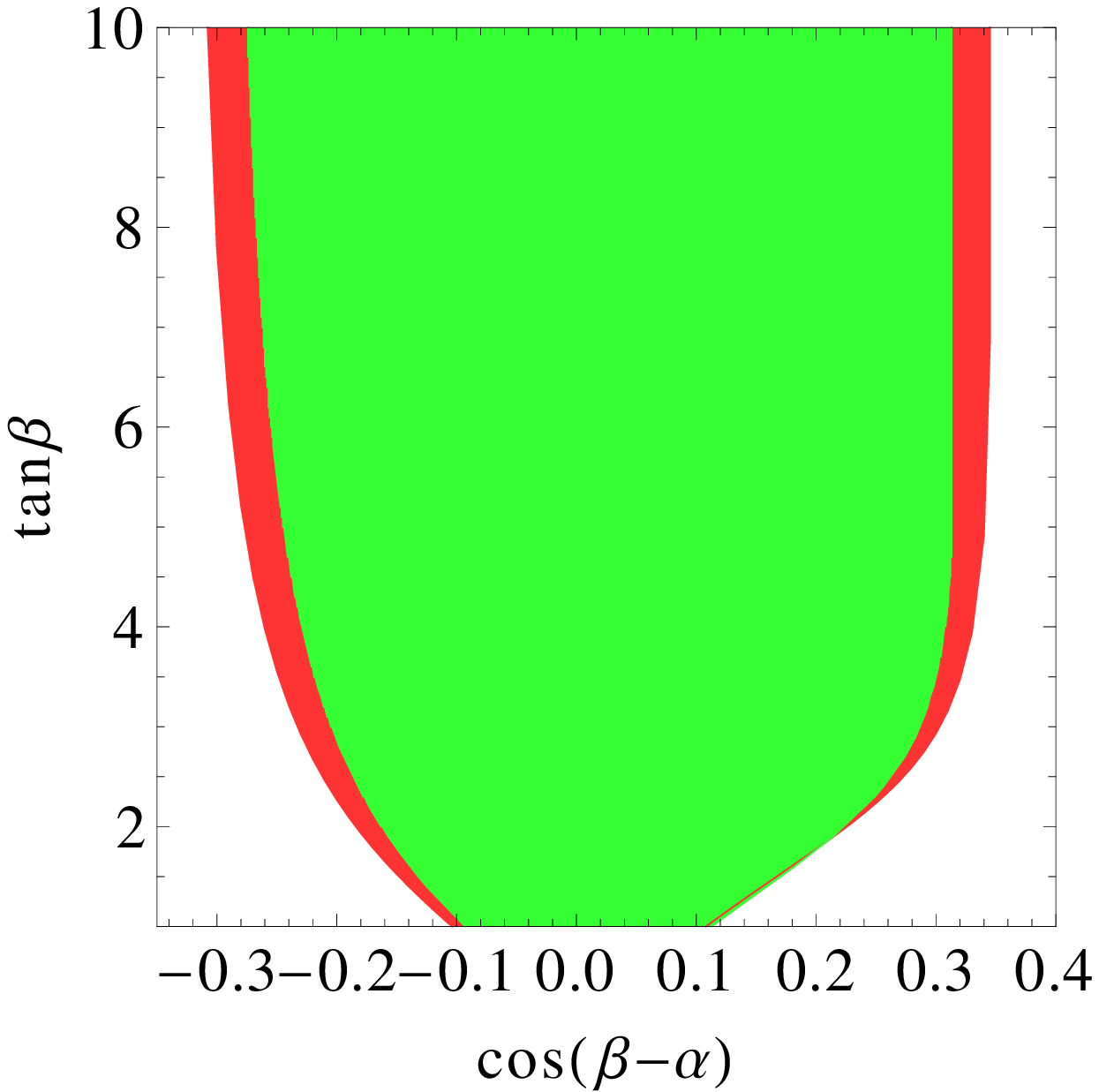}
}
\subfigure[]{
      \includegraphics[width=0.36\textwidth,angle=0,clip]{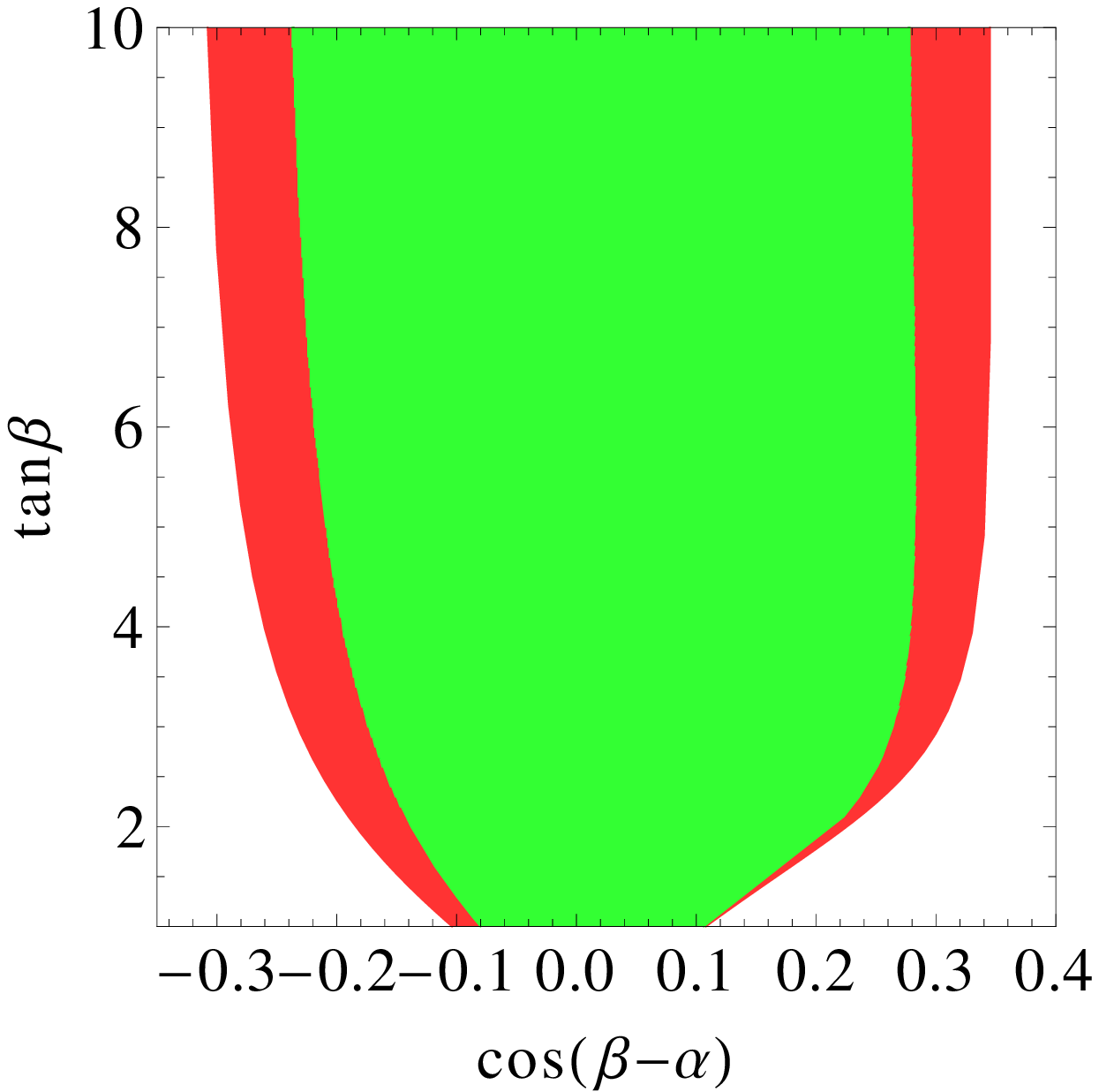}
}
\caption{Allowed regions in the $(\cos(\beta-\alpha),\tan\beta)$ plane  where the 
stronger constraint is provided by the non-observation of the heavy Higgs $H_3$ (light gray or green) or 
by the precision measurement of the SM Higgs $H_2$ (dark gray or red) at 95\% confidence 
level for an integrated luminosity of  300 fb$^{-1}$  
in the type-I N2HDM for (a) M$_{H_3}$ = 200 GeV and (b) M$_{H_3}$ = 600 GeV, obtained by performing  a $\chi^2$ analysis.
}
\label{mh3:200}
\end{figure}


\section{Conclusions}

Studies of Higgs properties at the LHC are concentrated in two areas.   There is intense study of the properties of the 126 GeV state, and there is a search for additional Higgs scalars, generally at larger mass.    An important question concerns the relative importance of these two areas in constraining extended Higgs sectors.

In the conventional 2HDM (either type-I or type-II), the answer is that, except for a small part of parameter space, the study of the properties of the 126 GeV Higgs is more useful\cite{Craig:2013hca,Chen:2013rba}.    However, the story might be different in the simplest extension of the 2HDM, a model with a real singlet.   In this article, we have shown that in type-I models, searches for heavy Higgs scalars will be more important for most of parameter-space, whereas in type-II models, study of the physics of the 126 GeV state is more critical.

How robust are these results?   We have assumed an $S \rightarrow -S$ symmetry.   If relaxed, then one could add terms such as $\lambda_{9ij} \sigma \Phi^\dagger_i\Phi_j S +$  h.c.   where $\sigma$ is an arbitrary mass scale.   These will affect all of the couplings and masses, and would result in a much more complicated analysis.   Of course, if $\lambda_9\sigma$ is small, the conclusions are unaffected.   Since the additional parameters are arbitrary, we will not discuss them further.

Most analyses of the capabilities of future facilities, such as the high luminosity  LHC and the ILC, focus on precision measurements of the Higgs properties.   Our result here shows the potential importance of heavy Higgs searches, even with reduced couplings to the SM particles, since in some very reasonable models, such searches can give stronger limits, or a discovery.

\vskip 3cm

\acknowledgments

We would like to thank Sally Dawson for helpful discussions.   The work of C-Y.C.is supported by the United States Department of Energy under Grant DE-AC02-98CH10886 and the work of M.F. and M.S. is supported by the National Science Foundation under Grant NSF-PHY-1068008.
The work of M.F. was supported by the JSA Initiatives Fund Program, an annual commitment from the JSA owners, SURA and CSC/ATD, to support programs, initiatives, and activities that further the scientific outreach, and promote the science, education and technology missions 
of Jefferson Lab and benefit the Lab user community. The opinions and conclusions expressed herein are those of the authors, and do not represent the National Science Foundation.
\newpage
\vskip -.5in
\input{tab-bosons}
\input{tab-fermions}

\end{document}

%% file: tab-bosons.tex
\begin{table}[tp]
\renewcommand{\arraystretch}{1.4}
\caption{Measured Higgs Signal Strengths}
\centering
\begin{tabular}{|c|c|c|}
\hline
 Decay          & Production & Measured Signal Strength $R^{meas}$ \\ \hline
{$\gamma \gamma$} 
                & ggF        & $1.6^{+ 0.3 +0.3}_{-0.3-0.2}$, [ATLAS] \cite{atlas-13012}\\ 
                & VBF        & $1.7^{+ 0.8 +0.5}_{-0.8-0.4}$  [ATLAS]\cite{atlas-13012}\\ 
                & Vh        & $1.8^{+ 1.5 +0.3}_{-1.3-0.3}$  [ATLAS]\cite{atlas-13012}\\ 
                & inclusive  & $1.65^{+ 0.24 +0.25}_{-0.24-0.18}$   [ATLAS]\cite{atlas-13012}\\ 
                & ggF+tth    & $0.52 \pm 0.5$   [CMS]\cite{cms-13001}\\ 
                & VBF+Vh     & $1.48^{+1.24}_{-1.07}$  [CMS]\cite{cms-13001}\\ 
                & inclusive  & $0.78^{+0.28}_{-0.26}$      [CMS]\cite{cms-13001}\\
                & ggF        & $6.1^{+3.3}_{-3.2}$  [Tevatron]\cite{hcp:Enari}\\ \hline
%
{$W W$}             
                & ggF        & $0.82 \pm 0.36$          [ATLAS] \cite{atlas-13030}\\ 
                & VBF+Vh     & $1.66 \pm 0.79$    [ATLAS]\cite{atlas-13030}\\ 
                & inclusive  & $1.01 \pm 0.31$   [ATLAS]\cite{atlas-13030}\\ 
                & ggF        & $0.76 \pm 0.21$        [CMS]\cite{cms-13003}\\ 
                & ggF        & $0.8^{+0.9}_{-0.8}$    [Tevatron]\cite{hcp:Enari}\\ \hline
%
{$ZZ$}              
                & ggF        & $1.8^{+0.8}_{-0.5}$   [ATLAS] \cite{atlas-13013}\\ 
                & VBF+Vh     & $1.2^{+3.8}_{-1.4}$   [ATLAS]\cite{atlas-13013}\\ 
                & inclusive  & $1.5 \pm 0.4$         [ATLAS]\cite{atlas-13013}\\ 
                & ggF        & $0.9^{+0.5}_{-0.4}$   [CMS] \cite{cms-13002}\\ 
                & VBF+Vh     & $1.0^{+2.4}_{-2.3}$   [CMS]\cite{cms-13002}\\ 
                & inclusive  & $0.91^{+0.30}_{-0.24}$ [CMS]\cite{cms-13002}\\ \hline
\end{tabular}

\vspace{-1ex}
\label{tab:models1}
\end{table}

%% file: tab-fermions.tex
\begin{table}[tp]
\renewcommand{\arraystretch}{1.4}
\caption{Measured Higgs Signal Strengths}
\centering
\begin{tabular}{|c|c|c|}
\hline
 Decay          & Production & Measured Signal Strength $R^{meas}$ \\ \hline
%
{$b\bar{b}$}      
                & Vh         & $-0.4 \pm 1.0$    [ATLAS] \cite{atlas-170}\\ 
                & Vh         & $1.3^{+0.7}_{-0.6}$      [CMS]\cite{cms-044}\\ 
                & Vh         & $1.56^{+0.72}_{-0.73}$   [Tevatron]\cite{hcp:Enari}\\ \hline
{$\tau^+ \tau^-$}
                & ggF        & $2.4 \pm 1.5$          [ATLAS]\cite{atlas-160}\\ 
                & VBF        & $-0.4 \pm 1.5$         [ATLAS]\cite{atlas-160}\\ 
                & inclusive  & $0.8 \pm 0.7$          [ATLAS]\cite{atlas-170}\\ 
                & ggF        & $0.73 \pm 0.50$        [CMS]\cite{cms-13004}\\ 
                & VBF        & $1.37^{+0.56}_{-0.58}$  [CMS]\cite{cms-13004}\\ 
                & Vh         & $0.75^{+1.44}_{-1.40}$    [CMS]\cite{cms-13004}\\ 
                & inclusive  & $1.1 \pm 0.4$        [CMS]\cite{cms-13004}\\ 
                & ggF        & $2.1^{+2.2}_{-1.9}$    [Tevatron]\cite{hcp:Enari}\\ \hline                 
\end{tabular}

\vspace{-1ex}
\label{tab:models2}
\end{table}